\documentclass[prl,showpacs,superscriptaddress,amsfonts,amsmath,floatfix,twocolumn]{revtex4-1}

\usepackage{graphicx}
\usepackage{hyperref} 
\usepackage{verbatim} 
\usepackage{url}
\usepackage{epstopdf}
\usepackage{natbib}
\usepackage{color}
\usepackage{dsfont}
\usepackage[utf8]{inputenc}
\usepackage{float}
\usepackage{amsmath}
\usepackage{comment}
\usepackage{breqn}
\usepackage{amssymb}
\usepackage{bbold}
\usepackage{siunitx}

\renewcommand\vec[1]{\ensuremath\boldsymbol{#1}}
\newcommand{\affA}{Department of Physics and Astronomy, Aarhus University, Ny Munkegade 120, Denmark}
\newcommand{\affB}{Instituto de F{\'i}sica da UFRGS, Av. Bento Gon{\c c}alves 9500, Porto Alegre, RS, Brazil}
\newcommand{\affC}{Aarhus Institute of Advanced Studies, Aarhus University, DK-8000 Aarhus C, Denmark}

\begin{document}
\title{Realizing time crystals in discrete quantum few-body systems}

\author{R. E. Barfknecht}
\email{rafael@phys.au.dk}
\email{rafael.barfknecht@ufrgs.br}
\affiliation{\affB}
\affiliation{\affA}

\author{S. E. Rasmussen}
\email{stig.elkjaer.rasmussen@post.au.dk}
\affiliation{\affA}

\author{A. Foerster}
\email{angela@if.ufrgs.br}
\affiliation{\affB}

\author{N. T. Zinner}
\email{zinner@phys.au.dk}
\affiliation{\affA}
\affiliation{\affC}

\date{\today}
\begin{abstract} 
\noindent The exotic phenomenon of time translation symmetry breaking under periodic driving - the {\it time crystal} - has been shown to occur in many-body systems even in clean setups where disorder is absent. In this work, we propose the realization of time-crystals in few-body systems, both in the context of trapped cold atoms with strong interactions and of a circuit of superconducting qubits. We show how these two models can be treated in a fairly similar way by adopting an effective spin chain description, to which we apply a simple driving protocol. We focus on the response of the magnetization in the presence of imperfect pulses and interactions, and show how the results can be interpreted, in the cold atomic case, in the context of experiments with trapped bosons and fermions. Furthermore, we provide a set of realistic parameters for the implementation of the superconducting circuit.
\end{abstract}


\maketitle

\paragraph*{Introduction.} Recent developments in theory and experiments with time-dependent quantum mechanical set-ups have consolidated the concept of the {\it discrete time crystal}, a system that presents spontaneous breaking of time-translation symmetry. 
The original proposals, both in the quantum \cite{wilczek} and classical \cite{wilczek_classical} regimes, suggested the possibility of a system exhibiting a periodic dynamical behavior in its lowest energy state \cite{li_time_crystals}. This possibility seems to have been ruled out by subsequent discussions \cite{bruno_tc_com1,bruno_tc_com2}, including no-go theorems for a broad class of systems \cite{bruno_tc_nogo,watanabe}. 

Surprisingly, it was later shown that systems in the presence of periodic driving, generally described by Floquet theory, can indeed self-organize and present a subharmonic response in the observables \cite{sondhi1,else,sacha_review}. The phase which exhibits the features of spatiotemporal order now recognized in time crystals was also classified as the $\pi$ spin glass \cite{sondhi2}, and since then a precise definition has been put forward \cite{sondhi3}. The properties of these systems have been studied in radically different configurations, such as atoms bouncing off an oscillating mirror \cite{sacha_modeling} or spin chains in the presence of disorder and many-body localization \cite{yao}. The latter proved to be an ideal starting point for experiments and resulted in the first two observations of time crystals. While these experiments dealt with two rather distinct arrangements (one exploring nitrogen vacancies in diamonds \cite{choi2017} and the other a chain of trapped ions \cite{zhang2017}), both had as a major feature the presence of disorder.

Other studies have shown, however, that disorder is not a crucial requirement for the realization of discrete time crystals \cite{sondhi4}. Cold atomic systems, for instance, can sustain a time crystal phase even in a ``clean" set-up where disorder is absent \cite{clean_time_crystals}. Experimentally, time crystal phases have been observed in ordered spin systems with nuclear magnetic resonance techniques \cite{pal,rovny}. In these studies, the presence of interactions between atoms or spins is the decisive factor leading up to self-organization and symmetry breaking. The manifestation of quasi-crystalline order and its transition to a time crystal has been observed with Bose-Einstein condensates under the action of periodically-driven magnetic fields \citep{autti}.

\begin{figure}
\centering
\includegraphics[width=0.85\columnwidth]{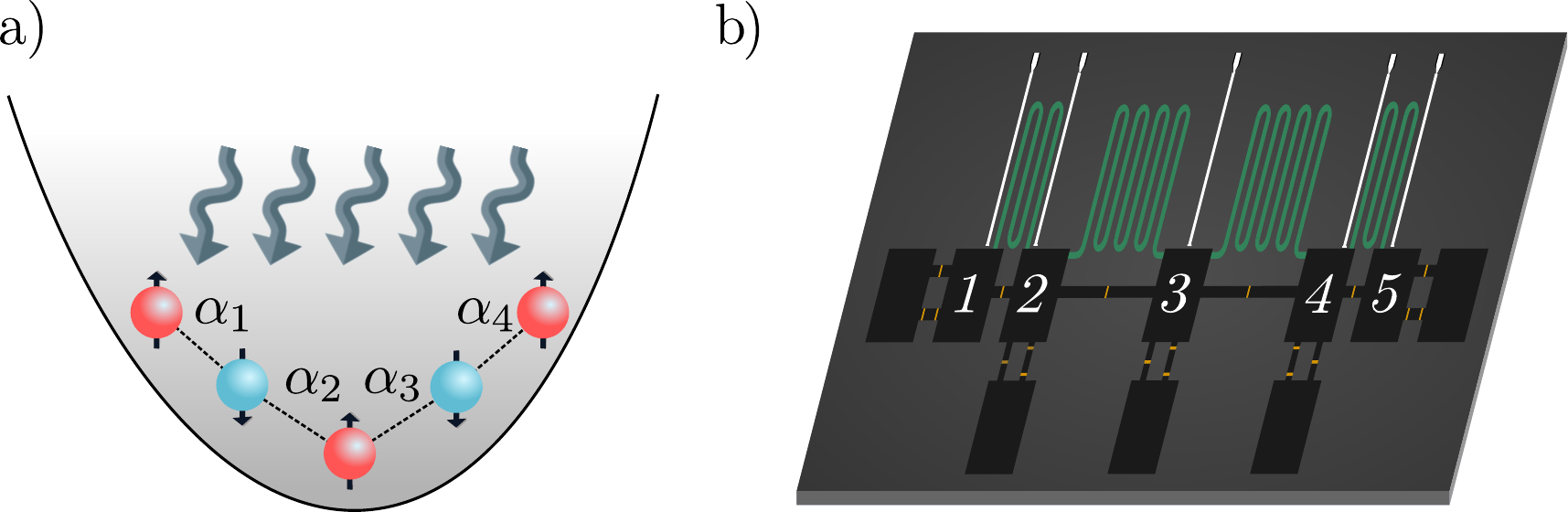}
\caption{Schematic depiction of the two few-body systems under consideration: a) an ensemble of cold atoms in a harmonic trap, where the strong interactions allow for a mapping of the system to a spin chain with position-dependent exchange coefficients, and b) a superconducting circuit consisting of five superconducting islands, each corresponds to a spin (see the Supplemental Material \cite{sm} for a detailed representation of the circuit in a chip). The driving protocol can be realized by applying spin-flip pulses (represented by the gray arrows on the left panel). In the case of the superconducting circuit, the pulses can be introduced as Rabi oscillations acting on each island through the white control lines.}
\label{fig1}
\end{figure}

These developments raise the question of how simple a system that exhibits a time crystal phase can be. To address this matter, we present a proposal for the realization of time crystals in few-body systems, which can be realized in two ways (see Fig.~\ref{fig1} for a schematic depiction of each case). The first one involves a small ensemble of cold trapped atoms, where the interactions between different components can be tuned by means of Feshbach or confinement induced resonances \cite{olshanii}. In the limit of strong interactions, the system can behave as a spin chain, where the exchange coefficients are determined by the shape of the trapping potential \cite{deuretz2,artem1}. Periodically driving the system with spin-flip pulses \cite{koutentakis} then results in a response in the magnetization which depends highly on the choice of interaction parameters. Moreover, we show that realizing the system with fermionic or bosonic atoms - the latter case assuming that interactions between identical particles can be tuned - yields very distinct results. 

The second realization is based on a circuit consisting of five superconducting islands coupled via Josephson junctions and inductors \cite{martinis1,nori1,devoret_outlook,gu}. Superconducting circuits can be used as platforms for simulations in atomic physics and quantum optics \cite{buluta,nori2}. Under certain conditions, such a system can also be interpreted as a spin chain \cite{heras,martinis2}; more importantly, by detuning the frequencies of the spins in our model and employing the rotating wave approximation (RWA), we get a Hamiltonian described only by Ising couplings. By then applying a strong external field to the islands for a short period of time, Rabi oscillations can be induced to each spin, causing them to flip. 

\paragraph*{System description}
We consider a general XXZ spin chain described by the following Hamiltonian:

\begin{eqnarray}\label{hamiltonian}
H&=&\sum_{i=1}^{N-1}\left(\eta_i^0+\eta_i^x\sigma_i^x\sigma_{i+1}^x+\eta_i^y\sigma_{i}^y\sigma_{i+1}^y+\eta_i^z\sigma_i^z\sigma_{i+1}^z\right)\nonumber \\
&-&\frac{1}{2}\sum_i^N \Omega_i \sigma_i^z
\end{eqnarray}
where we take a set of inhomogeneous exchange coefficients $\eta_i$. The values of these coefficients will be determined next, according to the model under consideration. We also include a site-dependent constant $\eta_i^0$ and the possibility of an inhomogeneous external field acting with a frequency $\Omega_i$. Contrary to some theoretical approaches \cite{ho_dipolar,russomano_lmg} and paradigmatic experimental realizations of time crystals \cite{choi2017}, our model has no need for long-range interactions.
We focus on a small system of $N=5$ spins and our protocol for the external driving is fairly simple: we choose an initial antiferromagnetic state, such as $\vert \psi(0) \rangle=\vert \uparrow\downarrow\uparrow\downarrow\uparrow\rangle$, which is not an eigenstate of the spin chain under any non-trivial parameter configuration. We then proceed to realize a sequence of spin-flip operations (with period $T_D$) at each site, rotating all spins by an angle $\theta$. We consider initially an instantaneous rotation, but we extend our results in superconducting circuits to the case of finite-time rotations, which are more realistic from an experimental viewpoint. By keeping track of the time-evolution of the magnetization $m(t)=\langle \psi(t)\vert \sum_{i=1}^N \sigma_i^z\vert \psi(t)\rangle$, we are able to register the response of the system with respect to the external driving. For perfect ($\theta=\pi$) pulses, we can expect the magnetization to have a trivial periodicity of $T_m=2T_D$. However, for imperfect rotations described by $\theta=\pi-\epsilon$, we may obtain a different response, which is strongly determined by the presence and nature of the interactions between the spins.

\paragraph*{Two-component trapped cold gases.} 
Our first application of the protocol described above is an interacting system of trapped cold atoms. Up to this point, most proposals for the realization of time-crystals with ultracold atoms involve the many-body problem of interacting identical bosons in the presence of a periodic Hamiltonian \cite{sacha_exp,sacha_phase_transition}, which is generally described by the Gross-Pitaevskii equation. In the present case, we focus on the few-body problem of bosonic atoms with two internal components, which we label as the pseudospin states $\vert \uparrow \rangle$ and $\vert \downarrow\rangle$. We consider contact interactions given by $g\sum_{i<j}\delta(x_i-x_j)$ for atoms in different internal states and $\kappa g \sum_{i<j}\delta(x_i-x_j)$ for atoms in the same internal state. All atoms are confined by an effectively one-dimensional harmonic trap described by $V(x)=\frac{1}{2}m \omega^2 x^2$. For simplicity, we assume the atoms to have the same mass $m=1$, and define the trapping frequency as $\omega=1$. 

In the limit of strong interactions ($g\gg 1$), it has been established that this system can be described, up to linear order in $1/g$, by an effective XXZ spin chain \cite{deuretz2,artem2} (see Supplemental Material \cite{sm}), which can be obtained from Hamiltonian \eqref{hamiltonian}. In this description, we make the substitutions $\eta_i^0=-\frac{1}{2}\frac{\alpha_i}{g}\left(1+\frac{2}{\kappa}\right)$, $\eta_i^x=\eta_i^y=\frac{1}{2}\frac{\alpha_i}{g}$ and $\eta_i^z=\frac{1}{2}\frac{\alpha_i}{g}\left(1-\frac{2}{\kappa}\right)$, where $g$ and $\kappa$ have the meaning defined above, and $\alpha_i$ are position-dependent exchange coefficients, which are determined solely by the geometry of the trap. These coefficients have been calculated for traps of different shapes and systems of up to $N\approx 30$ \cite{conan,deuretz_mdist}. Particularly, for the case of $N=5$ in a harmonic trap with trapping frequency $\omega=1$, we have $\alpha_1=\alpha_4\approx 2.16612$ and $\alpha_2=\alpha_3\approx 3.17738$, where the symmetry of the coefficients in guaranteed by the parity invariance of the trap. With this mapping (and considering $\Omega_i=0$), Hamiltonian \eqref{hamiltonian} can now be written as $H=-\frac{1}{2}\sum_{i}^{N-1}\frac{\alpha_i}{g}\left[\frac{1}{2}\left(1-\vec{\sigma}_i\cdot \vec{\sigma}_{i+1}\right)+\frac{1}{\kappa}\left(1+\sigma_i^z\sigma_{i+1}^z\right)\right]$, where the magnitude of the Ising term on the right side is now directly determined by the value of the intraspecies interaction parameter $\kappa$. While this Hamiltonian describes a system of strongly interacting bosons (due to the presence of interactions between identical components), we can reproduce a fermionic system by taking the limit $\kappa\rightarrow \infty$. Then, we obtain $H=-\sum_{i=1}^{N-1}\frac{\alpha_i}{g}(1-P_{i,i+1})$, where $P_{i,i+1}=\frac{1}{2}(1+\vec{\sigma}_i\cdot \vec{\sigma}_{i+1})$ is the permutation operator that exchanges neighboring spins. We now focus on initializing the system in the state described above and calculating the time-evolution of the magnetization under the periodic action of the spin-flip operator $O=\exp{(-i\frac{\theta}{2}\sum_{i=1}^{N}\sigma_i^x)}$.

\begin{figure}
\centering
\includegraphics[width=0.85\columnwidth]{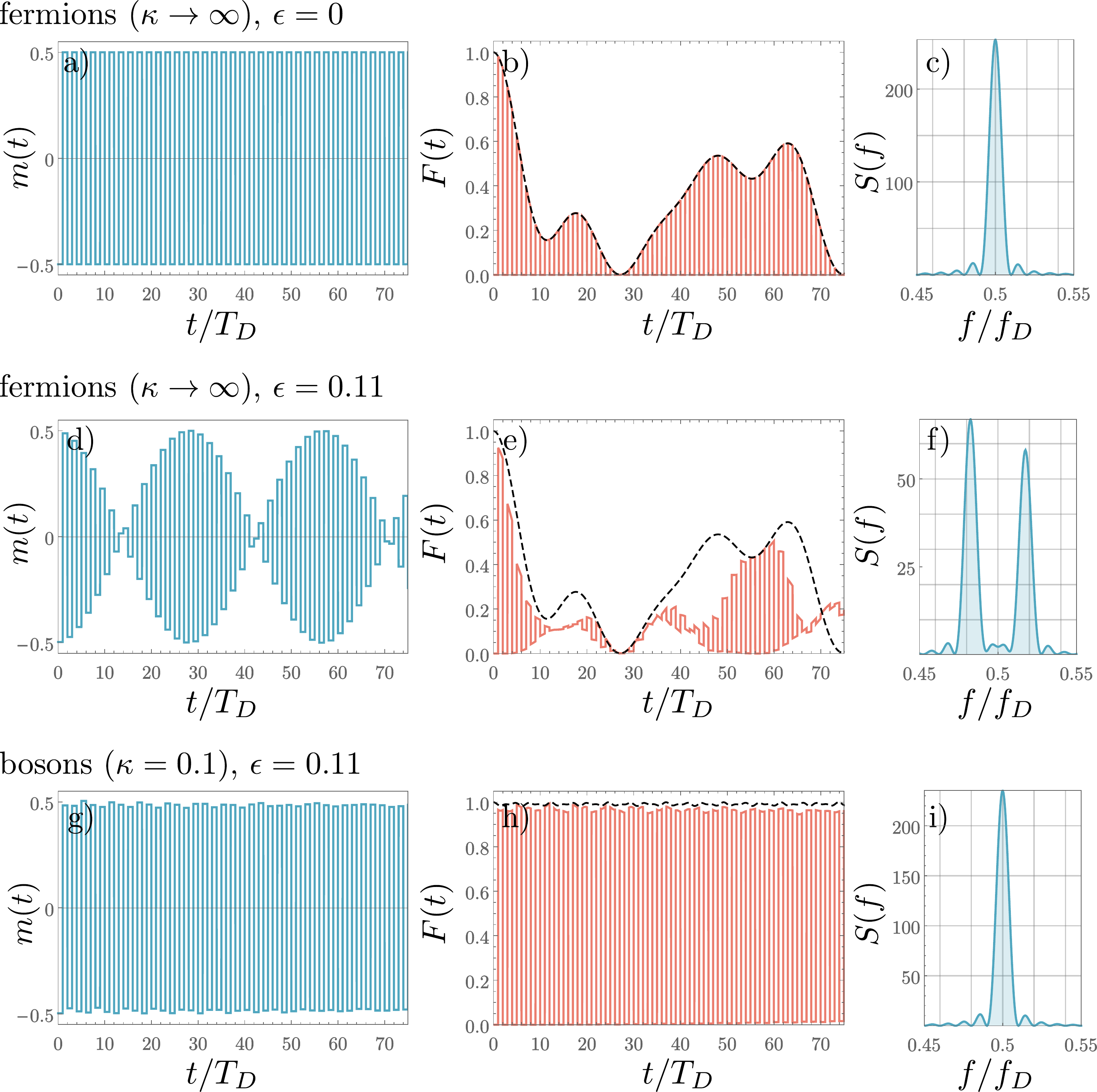}
\caption{Time evolution of observables in a system of $N=5$ harmonically trapped atoms. The parameter $\kappa$ defines the atomic species (fermionic or bosonic) and the value of $\epsilon$ determines the presence of imperfections in the driving (for $\epsilon=0$ we have a perfect $\pi$-pulse). Panels a), d) and g) show the time evolution of the magnetization, while panels b), e) and h) present the analogous results for the overlap with the initial state $F(t)$. In these figures, the black dashed curves show the results for $F(t)$ in the absence of driving. Panels c), f) and i) show the spectral density obtained through the Fourier transform of $m(t)$.}
\label{fig2}
\end{figure}

In Fig.\ref{fig2} a), b) and c) we show, respectively, the results for the time-evolution of the magnetization $m(t)$, the overlap probability of the wave function with the initial state $F(t)=\vert\langle \psi(0)\vert \psi(t)\rangle\vert^2$ and the spectral density $S(f)=|\hat{m}(f)|^2$, where $m(f)$ is the Fourier transform of the magnetization, given by $S(f)=\vert \int dt\,e^{-2\pi i f t}m(t)\vert ^2$. We initially assume a periodic pulse that rotates the spins by an angle of $\theta=\pi$ at times $t=nT_D$ with $n$ being an integer. We find that the magnetization is accordingly switched between +0.5 and -0.5 with a period twice as large as the driving, which results in a peak in $f=f_D/2$ where $f_D$ is the driving frequency. While this quantity only registers the global behavior of the system, the overlap with the initial state $F(t)$ describes its underlying spin dynamics. When all spin are rotated by $\pi$ with respect to the initial state, we have $F(t)=0$. In the remaining times, we observe that the time-evolution of the spin distribution is described by the exact results in the absence of periodic driving, which is shown as as black dashed curve. 

If we consider an imperfect pulse defined by the angle $\theta=\pi-\epsilon$, we observe two different results. In the case of fermions (Figs. \ref{fig2} d), e) and f)), the magnetization now exhibits a beating pattern that destroys the subharmonic peak at $f=f_D/2$. The overlap $F(t)$ is no longer described by the result in the absence of driving. For bosons (Figs. \ref{fig2} g), h) and i)), on the other hand, the presence of a dominating interaction between identical spins - defined by the small value of $\kappa$ - locks back the magnetization response peak at $f=f_D/2$, even for $\epsilon\neq 0$. This robustness of the system's response in the presence of imperfect pulses is a defining feature of a time crystal phase \cite{yao}. In fact, here we find that it arises under fairly simple conditions, without the need of switching interactions on and off as part of the driving protocol.  Moreover, it presents the possibility of studying the ``melting" of time crystals as interactions are dynamically changed. This could be implemented, for instance, by taking a bosonic system and tuning $\kappa$ from small to large. It is important to point out that experiments with only a small number of trapped cold atoms are available and that interactions can indeed be manipulated over wide ranges \cite{jochim3}. Moreover, the driving protocol employed here can be also used in the case of systems with more than two internal components, as long as Ising-type interactions are dominant in the Hamiltonian. By periodically switching between different pseudospin states, it is possible to expect a fractional response frequency given by $f=f_D/\nu$ where $\nu$ is the number of internal states available. Multi-component cold atomic gases, such as fermionic systems with SU(N) symmetry, have been theoretically explored \cite{xiaoling,laird}, and can be realized in the lab \cite{fallani,fallani_hall}. A recent proposal for realizing time-crystals in SU(N) systems explores the ladder of internal states as a synthetic dimension \cite{clean_time_crystals}. In a more extreme example, systems where the response frequency is much smaller than the driving frequency have been obtained in the framework of atoms bouncing off an oscillating mirror \cite{sacha_exp}. 

\begin{figure}[H]
	\centering
	\includegraphics[width=0.8\columnwidth]{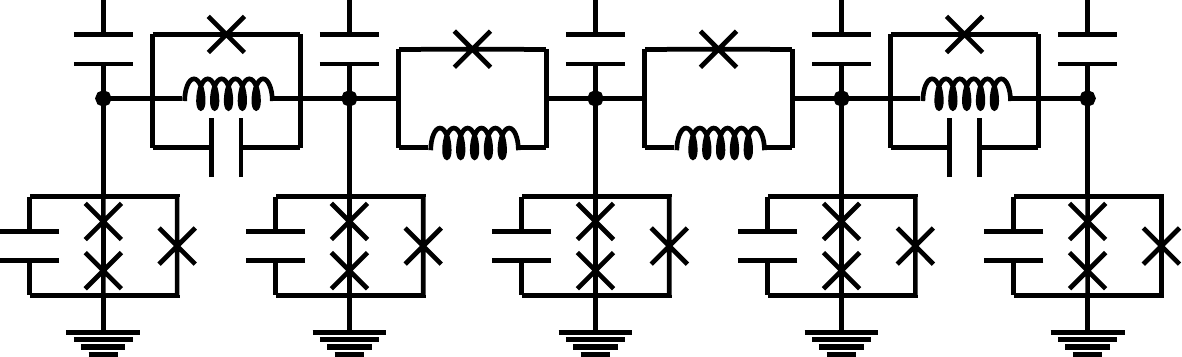}
	\caption{Circuit diagram for a superconducting circuit of $N=5$ qubits. The C-shunted flux qubits are connected by Josephson junctions and inductors, and the outer qubits are also connected by capacitors. An external driving field is applied to each node.}
	\label{fig3}
\end{figure}

\paragraph*{Superconducting circuits.} 
As a second implementation, we apply the formalism described above to the case of a superconducting circuit. The circuit consists of five C-shunted flux qubits \cite{Yan2016}, which are particularly interesting for experimental realization due to their long decoherence time, but other types of superconducting qubits can be used as well \cite{girvin1}. The qubits are connected pairwise with Josephson junctions and inductors, and the outer islands are also connected through capacitors (see Fig. \ref{fig3} for a lumped circuit model). An external driving field is further applied to each node, in order to drive the rotation of spins.

By applying Devoret's quantum treatment of electromagnetic circuits \cite{Devoret1997,Devoret2017}, we find the Hamiltonian of the system, which can again be taken as a particular case of Eq. \ref{hamiltonian} (see the Supplemental Material \cite{sm} for details on this derivation). Due to the anharmonicity of the Josephson junctions we can truncate our description into the two lowest energy levels, which leaves us with an effective spin model. By requiring a detuning of the spin frequencies, we use the RWA to remove all interaction terms but the Ising couplings $J_i^z\sigma_{i}^z \sigma_{i+1}^z$, where $J_i^z=\eta_i^z$ in Eq. \eqref{hamiltonian}. In order to realize the rotations between internal states we apply a strong driving to the the circuit for a short period of time. In our truncated spin model, using ladder operators, the driving can be described by
\begin{equation}
H_d = i A \sum_{i=1}^{N} \cos(\omega_i t)( \sigma_i^+e^{i\Omega_i t} - \sigma_i^-e^{-i\Omega_i t}),
\end{equation}
where $A$ is the amplitude of the external fields, $\omega_i$ is the driving frequency and $\Omega_i$ is the qubit frequency. For the driving to cause Rabi oscillations between $\vert \uparrow \rangle$ and $\vert \downarrow \rangle$ we must require $\omega_i = \Omega_i$ and $A \gg J_i$. The driving period is thus given by $\Delta t = (\pi-\epsilon)/A$, which, for large $A$, is a very short time. 
Choosing realistic variables for the circuit, we obtain a set of experimentally achievable parameters for our spin model. In Table \ref{table1} we show the relevant values for the frequencies $\Omega_i$ and the exchange coefficients $J^z_i$ used in the interacting cases. 

\begin{table}
	\centering
	\begin{tabular*}{0.45\textwidth}{@{\extracolsep{\fill}} lccc}
		\hline
		\multicolumn{4}{c}{}\\
		Site $i$ & 1 & 2 & 3 \\
		\hline 
		$\Omega_i/2\pi\,\si{\GHz}$ & $ \num{17.0 +- 0.048} $ & $ \num{35.6 +- 0.21} $ & $ \num{43.361 +- 0.048} $ \\ 
		$J^z_{i}/2\pi\,\si{\MHz}$ & $ \num{168.9 +- 1.1}$ & $ \num{-29.07 +- 0.18} $ & - \\ 
		\hline
	\end{tabular*}
\caption{Spin model parameters used in our simulations. For the non-interacting cases, we make $J_i^z=0$. The numbers between parenthesis are the uncertainty related to the numerical values of the circuit parameters (see \cite{sm} for details).}
\label{table1}
\end{table}

\begin{figure}
\centering
\includegraphics[width=0.85\columnwidth]{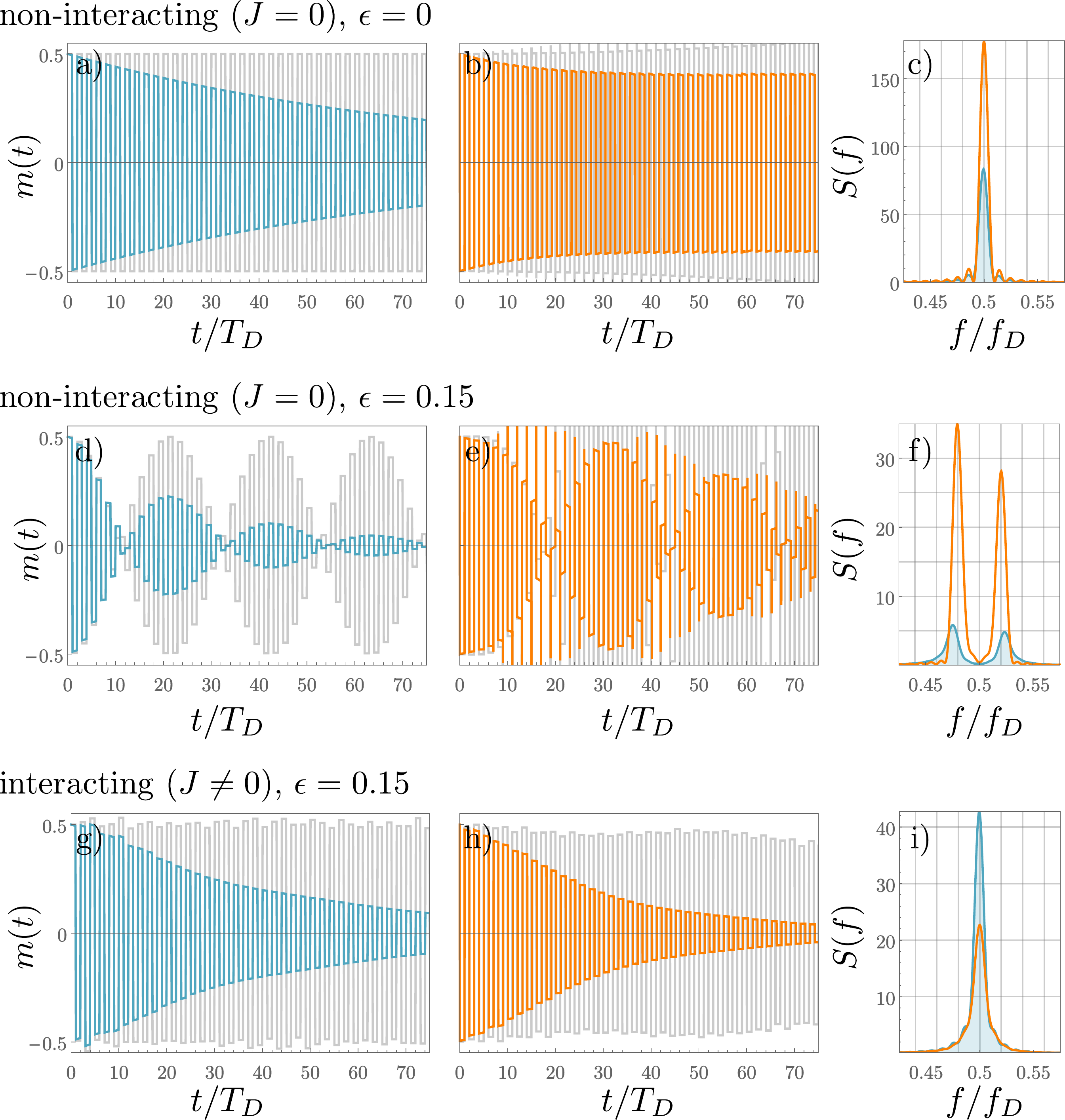}
\caption{Time evolution of the magnetization in circuits with a), d) and g) ideal driving and b), e) and h) inhomogeneous driving. In all plots, the colored curves show the results with an added relaxation noise of $\zeta=0.05$, while the gray background curves depict the results without losses. The non-interacting cases assume $J_i^z=0$, while in the interacting cases the parameters are given by Table~\ref{table1}. Plots c), f) and i) show the spectral density of the magnetization, where the colors correspond to the cases of ideal (blue) and inhomogeneous (orange) driving shown on the left.}
\label{fig4}
\end{figure}

We then run simulations for the driving described in this section using QuTip \cite{QuTip2013}, both with and without dephasing and relaxation noise $\zeta$. The result of these simulations are presented in Fig. \ref{fig4}. 
It shows that the different driving in each qubit leads to an interference pattern in the magnetization which is not due to imperfections in the pulse. An interesting effect of this interference is that it can cause the absolute value of the magnetization to be larger than $0.5$ during the driving period. In this particular case, we find the interference to be constructive, but destructive interference patterns are also possible. In the presence of imperfect pulses ($\epsilon\neq 0$), we now have two sources of interference acting on the system - the imperfect pulse itself and the driving inhomogeneities. This again leads to a splitting of the half-frequency peak, both in the ideal and inhomogeneous case. It turns out, however, that the addition of a strong Ising coupling fixes both of these effects at once, without any need of fine tuning of the parameters. An important feature of our superconducting circuit model is that the interactions between qubits can be different as long as there is symmetry across the center. 

\paragraph*{Conclusions.} 

We have presented a proposal for the realization of discrete time crystals in few-body spin chains in the presence of external driving, which can be realized in two distinct ways. In the first case, we apply our formalism to a system consisting of strongly interacting harmonically trapped atoms. We show how a time crystal regime can arise under the action of a periodic spin-flip driving, provided that the intraspecies repulsion is smaller than the remaining interactions. This leads to the possibility of studying time crystallization in bosonic as opposed to fermionic systems, or even the dynamical ``melting" of the time-crystal as the intraspecies interactions are tuned from weak to strong. Our second realization shows how a similar result can be achieved in a superconducting circuit, where our parameters can be manipulated to achieve an effective spin chain description. In this case, we can introduce the effect of losses and inhomogeneous driving in addition to the imperfection in the pulses. The possibility of a periodic response which is robust against such imperfections and inhomogeneities is an interesting feature for the implementation of quantum devices based on superconducting circuits \cite{devoret_outlook}. We also present a set of parameters for our quantum model which arise from a choice of realistic parameters for the circuit model. 
We stress the fact that in neither of these approaches we have to introduce disorder, which is a common features in previous studies. Naturally, the results described in our study can be generalized to larger chains or systems with more internal degrees of freedom.

\begin{acknowledgments}
\paragraph*{Acknowledgments.}
The authors thank O. V. Marchukov and A. G. Volosniev for fruitful discussions on time crystals. The following agencies - Conselho Nacional de Desenvolvimento Científico e Tecnológico (CNPq), the Danish Council for Independent Research DFF Natural Sciences, the DFF Sapere Aude program and the Carlsberg Foundation Distinguished Fellowship program - are gratefully acknowledged for financial support.
\paragraph*{Added note.} During the completion of this manuscript we became aware of a related work by W. C. Yu {\it et al.} \cite{yu_time_crystal} which discusses similar ideas. Our results 
are consistent with that work where there is overlap. In contrast to that study, we show here that the results hold in the few-body limit, which is the situation for many of the current experimental setups.
\end{acknowledgments}

\bibliographystyle{ieeetr}
\bibliography{biblio}


\end{document}